\documentclass[prd,a4paper]{revtex4}
\usepackage{graphicx,amsthm,amsmath,latexsym,amssymb,amsfonts}
\usepackage{dcolumn,calrsfs,fancyhdr,array,stmaryrd,euscript}
\usepackage{bm}
\usepackage{color}
\usepackage{wasysym}

\begin{document}

\title{Anyonic glueballs from an effective-string model}

\author{Nicolas \surname{Boulanger} }
\email[E-mail: ]{nicolas.boulanger@umons.ac.be}
\affiliation{Groupe de M\'ecanique et Gravitation, Physique Th\'eorique et Math\'ematique, Universit\'{e} de Mons -- UMONS,
Place du Parc 20, 7000 Mons, Belgium}
\author{Fabien \surname{Buisseret}}
\email[E-mail: ]{fabien.buisseret@umons.ac.be}
\affiliation{Haute \' Ecole Louvain en Hainaut (HELHa), Chauss\'ee de Binche 159, 7000 Mons, Belgium} 
\affiliation{Service de Physique Nucl\'{e}aire et Subnucl\'eaire,
Universit\'{e} de Mons -- UMONS,
Place du Parc 20, 7000 Mons, Belgium}

\date{\today}

\begin{abstract}
Relying on an effective-string approach in which glueballs --- bound states of pure Yang-Mills theory --- are modelled by closed strings, we give arguments suggesting that anyonic glueballs, \textit{i.e.} glueballs with arbitrary spin, may 
exist in $(2+1)$-$\,$dimensional Yang-Mills theory. We then focus on the large$\,$-$N_c$ limit of $SU$($N_c$) Yang-Mills theory and show that our model leads to a mass spectrum in good agreement with lattice data in the scalar sector, while it predicts the masses and spins of anyonic glueball states. 
\end{abstract}



\maketitle

\section{Introduction}

The appearance of quantum states with arbitrary spin, called anyons, 
is a fascinating feature of 
quantum mechanics in $(2+1)$ dimensions \cite{Wilczek} that has been 
explored in a considerable 
amount of works: The interested reader may find useful references in 
\cite{Fort,Khare,Frohlich:1988qh}.  Actually, in $2+1$ dimensions 
the spin $s$ of a given state may be arbitrary because 
the Lorentz group $SO(2,1)$, as a group manifold, 
contains a non-contractible circle $S^1\,$ whose covering $\mathbb{R}$
covers it infinitely many times.
In the case of an Euclidean spacetime, the ``Lorentz" group $SO(3)\,$
is a compact, connected (albeit non-simply) manifold that admits at 
most two$\,$-valued unitary representations.  

It is known in field theory that coupling matter field to a 
three-dimensional vector gauge field with 
a Chern-Simons term leads to the appearance of states with 
fractional statistics \cite{semenoff}. 
The equivalent result is obtained within an $O(3)$ $\sigma$-model 
with Hopf term \cite{Wu}.
Note that a Chern-Simons term is not a necessary condition to produce 
anyons in field theory, as illustrated  by the following examples:

\begin{itemize}
\item Composite quantum states with arbitrary spin or arbitrary 
exchange statistics can be built from 
the genuine Abelian Higgs model without Chern-Simons 
term \cite{Ligu,Cherno};
\item Within an Abelian gauge theory with matter field denoted by 
$\Psi$ and $g^{2}$ a constant with 
dimension of mass, one defines the shifted 
connection $A^\theta_\mu=A_\mu+\frac{\theta}{g^2} F_\mu$ where
$F_\mu=\epsilon_{\mu\nu\rho}F^{\nu\rho}/2\,$.
The operator 
$\Psi(x) P\{ {\rm exp} (i \int^y_x dz^\mu A_\mu^\theta)\} \bar\Psi(y)$ then  
propagates 
an anyon with non trivial statistics related to the arbitrary real 
number $\theta$ \cite{Itzhaki}; 
\item The spectrum of closed Nambu-Goto strings in $2+1$ dimensions 
necessarily contains fractional
spin fields after light-cone quantization \cite{Mezi}.
\end{itemize}

More generally, it has to be stressed that the existence of 
fractional-spin 
fields in $(2+1)$-dimensional Minkowski spacetime arises from pure 
group theoretical arguments 
that are actually independent of the particular form of the action 
under consideration \cite{Barut:1965,Binegar:1981gv}. 
These arguments will be summarized in Sec. \ref{secgen}, 
while the case of closed Nambu-Goto strings, particularly important 
for our present work, will be discussed in Sec. \ref{secexam}.

The purpose of the present note is to investigate whether anyonic 
states exist or not in pure $(2+1)\,$-dimensional Yang-Mills theory. 
Such a problem has, to our knowledge, never been studied so far. If 
anyonic glueballs can be built, the next question is: What are their 
masses and spins? 
This problem can be addressed by resorting to a closed-string 
effective model of glueballs. 
The idea that Yang-Mills theory should be equivalent to some closed 
string theory at large $N_c$ actually
originates from 't Hooft and Veneziano's work on the large$\,$-$N_c$ 
limit of QCD \cite{'tHooft:1973jz,Vene}. 
It has indeed been known since then that any amplitude in 
large$\,$-$N_c$ Yang-Mills theory can be expressed as a sum 
over terms containing planar diagrams forming Riemann surfaces 
with various genus numbers, 
just as it is the case in closed string theory. 

From an effective model point of view, it is therefore tempting to 
assume that glueball dynamics has some stringy nonperturbative origin. 
The celebrated Isgur and Paton's flux tube model \cite{Isgu} is a 
first 
example of how, starting from a lattice-QCD-inspired approach, one is 
led to the conclusion that glueballs -- or at least some of them -- 
may be described by closed strings. Closed effective strings are 
often referred to as closed flux tubes since they are seen as 
particular configurations of the chromoelectric field whose dynamics 
is expected 
to be that of a closed string. Interestingly, lattice computations 
regarding Yang-Mills theory have given some support to this picture. 
The interested reader may find in \cite{John,Teper4} a discussion of 
the agreement between a string model of glueballs and the lattice data of \cite{Teper}. 

The effective model we use, inspired in particular by that of 
Ref. \cite{John}, is presented in Sec. \ref{secmodel} and numerical 
results are obtained in Sec. \ref{secresu}. Concluding comments are 
finally given in Sec. \ref{secconclu}.

\section{Relativistic anyons}\label{secgen}

Since the seminal works of Wigner and Bargmann \cite{Wigner:1939cj,Bargmann:1948ck}, 
it has been known that the elementary particles in 
Minkowski spacetime of dimension $D$ 
are associated with the unitary irreducible representations (UIRs) 
of the spacetime isometry 
group $ISO(D-1,1)\,$, the latter group being the semi-direct product 
of the Lorentz group $SO(D-1,1)$ with  the translation group 
${T}_{D}\,$. 
In $2+1$ dimensions, the Poincar\'e group therefore is 
$ISO(2,1)\cong SO(2,1)\ltimes {T}_{3}\,$. In this Section we first 
show how to build the UIRs of the $(2+1)-$dimensional Lorentz group 
$SO$(2,1); then we extend the discussion to the Poincar\'e group 
$ISO$(2,1).

\subsection{The case of SO(2,1)}

Let $L_{ab}=-L_{ba}$ be the generators of $SO(2,1)$. 
Then $so(2,1)$, the Lorentz algebra in $2+1$ dimensions, 
is presented by
\begin{equation}\label{alg0L}
[J^a ,J^b]= - i \varepsilon^{abc} J_c\;,
\end{equation}
where $J^a= \frac{1}{2}\, \varepsilon^{abc} L_{bc}$ and 
$\varepsilon^{012}=1\,$. Using the Minkowski metric in Cartesian 
coordinates $\eta={\rm diag}(-++)$ 
and bold fonts for 3-vectors, the scalar product of $\bm U$ and 
$\bm V$ reads 
$\bm U\cdot \bm V \equiv U^{a}\,\eta_{ab}V^{b}$ and 
the Casimir operator of $SO(2,1)$ is taken to be 
\begin{equation}\label{casimirL}
C_2[so(2,1)]:=-{\bm J}^2 = \tfrac{1}{2}\,L^{ab}L_{ab}\;.
\end{equation}
We use the notation $\vec u, \vec v\,\ldots\,$ for 2-vectors in the planes 
at fixed values of the Minkowskian coordinate $x^0\,$. 

An oscillator-based method for the 
classification of the UIRs of $SO$(2,1) 
was given in \cite{Barut:1965}, which we 
closely follow since it has the advantage of 
building up the UIRs of $SO$(3) in complete 
analogy, thereby unifying the treatments of 
the various real forms of $so(3,\mathbb{C})\,$.
It is of importance for us in view of drawing the reader's attention 
on the differences between both groups, the latter being actually at 
the basis of lattice QCD because of the Wick rotation leading to a 
Euclidean rather than hyperbolic spacetime. 

Defining, as usual, the ladder operators 
\begin{equation}
J^\pm=\frac{1}{\sqrt 2} ( - i J^2\pm J^1)\;, 
\end{equation}
yields
\begin{equation}
[ J^+ , J^- ] = \left\{ 
\begin{array}{ccc}
L_{12}  & \quad\mbox{for} & \quad so(2,1) \\
- L_{12} &  \quad\mbox{for} & \quad so(3)
\end{array}
\right.
\end{equation}
and 
\begin{equation}
[ L_{12} , J^\pm ] = \pm  J^\pm \;, 
\end{equation}
for both $so(2,1)$ and $so(3)$. The authors of \cite{Barut:1965} 
considered the complex algebra $so(3,\mathbb{C})\,$, thereby taking 
one and the same set of 
commutation relations for \emph{both} $so(2,1)$ and $so(3)\,$ 
(with $[ J^+ , J^- ] = L_{12}$) and distinguish the groups
$SO(2,1)$ and $SO(3)$ by different reality conditions on the 
corresponding parameters of infinitesimal transformations. In turn, 
these conditions and  the requirement of unitarity 
of irreducible representations for $SO(2,1)$ or $SO(3)\,$ give 
different reality conditions on the generators of the two groups. 
Effectively, this amounts to allowing real linear combinations 
of the non-compact generators $\{ L_{01} , L_{02} \}$ for $SO(2,1)$ 
and only purely imaginary linear combinations of them in the case of 
$SO(3)\,$, thereby euclideanizing $SO(2,1)$ to $SO(3)\,$, or stated 
equivalently, making $\{ L_{01} , L_{02} \}$ compact. 
One must have $(J^+)^\dagger=J^-$ for the rotation group $SO(3)$ and  
$(J^+)^\dagger=-J^-$ for the three-dimensional Lorentz group 
$SO(2,1)\,$.  
To summarise, in a unitary representation
\begin{eqnarray}
\begin{array}{ccc}
L_{12}^\dagger = L_{12}\;, \qquad  (J^+)^\dagger &= - J^- \quad & 
\;\;\mbox{for}\quad SO(2,1)\;,
\\
L_{12}^\dagger = L_{12}\;, \qquad   (J^+)^\dagger &\!\!\! = \; J^- 
\quad &\mbox{for}\quad SO(3)\;.  
\end{array}
\label{2.5}
\label{2.6}
\end{eqnarray}

Let $\xi=(\xi_{\alpha})_{\alpha=1,2} $ be a commuting real spinor 
of $SO(2,1)\,$ and consider the linear vector space spanned
by normalised vectors of the form
\begin{eqnarray}
|\Phi,m\rangle &=& {\cal N}_m\,\xi_1^a \xi_2^b = {\cal N}_m\, 
(\xi_1\xi_2)^{\Phi} \, (\xi_1/\xi_2)^{E_0 + m} \;, \nonumber\\
&&{\rm where}\ \Phi = \tfrac{1}{2}\,(a+b)\;, \quad E_0 + m = 
\tfrac{1}{2}\,(a - b)\;,\quad  (a,b) \in \mathbb{C}^2\;,\quad 
m\in\mathbb{Z}.
\label{4.1} 
\end{eqnarray}
The integer $m$ unambiguously labels the 
vectors once $\Phi$ and 
$E_0$ are specified. 
The inner product is defined by 
$\langle \Phi, m | \Phi,m'\rangle = \delta_{m,m'}$. 
In this representation the generators $J^{\pm}$ and $J^0$ are  
realised by the operators 
\begin{equation}
J^0 = 
\tfrac{1}{2}\,(\xi_1\frac{\partial}{\partial \xi_1} - 
\xi_2\frac{\partial}{\partial \xi_2 })\;,
\quad
J^+ = 
\tfrac{1}{\sqrt{2}}\, \xi_1\frac{\partial}{\partial \xi_2} \;,
\quad
J^- =
\tfrac{1}{\sqrt{2}}\, \xi_2\frac{\partial}{\partial \xi_1} \;.
\end{equation} 
These act on the basis vectors as
\begin{eqnarray}
\left\{ 
\begin{array}{ccc}
J^0 |\Phi , m \rangle  &=&  (E_0 + m) \,  |\Phi , m \rangle  \;,
\\
&   & \\
J^+ |\Phi , m \rangle  &=&  \tfrac{1}{\sqrt{2}}\, (\Phi - E_0 - m) 
\,({\cal N}_m/{\cal N}_{m+1})\, 
|\Phi , m+1 \rangle  \;,
\\
&    & \\
J^- |\Phi , m \rangle  &=&  \tfrac{1}{\sqrt{2}}\, (\Phi + E_0 + m) 
\,({\cal N}_m/{\cal N}_{m-1})\, 
|\Phi , m-1 \rangle  \;.
\end{array} \right. 
\label{4.2}
\end{eqnarray} 
One sees that $(E_0+m)$ is the eigenvalue of the generator of spatial 
rotations $J^0$, while a quick calculation shows that $\Phi (\Phi+1)$ 
is the eigenvalue of the quadratic Casimir 
\begin{equation}
C_2=2J^-J^++J^0(J^0+1)=2J^+J^-+J^0(J^0-1).
\end{equation}
Imposing the unitarity condition (\ref{2.5}) for the $SO(3)$ group 
leads to
\begin{eqnarray}
{\cal I}(E_0) &=& 0 \;, \\
{\left| \frac{{\cal N}_{m+1}}{{\cal N}_m}\,\right|}^2 
&=& \frac{\Phi^*-E_0-m}{m+E_0+\Phi+1} \;.
\end{eqnarray}
These recursion relations can be solved only if the values of $m$ are 
bounded both from above and from below. 
This corresponds to the well-known result that 
the UIRs of $SO(3)$ are finite-dimensional. 
The spectrum of the operator $J^0$ is then 
\begin{eqnarray}
\tfrac{1}{2}(a-b) = -\Phi, -\Phi + 1, \ldots , \Phi\;,
\end{eqnarray}
showing that $\Phi=s$ is the spin of the corresponding 
$SO(3)$ irreducible representation, and ${\cal R}E_0=0$.

For the non-compact group $SO(2,1)\,$, one derives from the 
unitarity condition (\ref{2.5}) that
\begin{eqnarray}
{\cal I}(E_0) &=& 0 \;, \\
{\left| \frac{{\cal N}_{m+1}}{{\cal N}_m}\,\right|}^2 
&=& \frac{m+E_0+\tfrac{1}{2}\,-( \Phi^*+\tfrac{1}{2} ) }{m+E_0+\tfrac{1}{2}+(\Phi+\tfrac{1}{2})} \;.
\label{4.4}
\end{eqnarray}
Keeping the notation of \cite{Barut:1965}, this gives the following 
UIRs :
\begin{eqnarray}
\begin{array}{ccc}
{\cal D}(C_2,E_0)\, & : & \, \quad C_2 < |E_0| (|E_0|-1)\quad {\rm and}\quad m\in \mathbb{Z} \; ,
\\
{\cal D}^+(\Phi)\, &:& \, \quad \Phi < 0\quad {\rm and} \quad m\in \mathbb{N}_{0} \;,
\\
{\cal D}^-(\Phi)\, &:& \, \quad \Phi < 0\quad {\rm and}\quad - m\in \mathbb{N}_{0} \;,
\\
{\cal D}(\Phi)\, &:& \, \quad \Phi=0  \quad {\rm and}\quad m=0 \;.
\end{array}
\label{4.5}
\end{eqnarray}

The UIRs ${\cal D}(C_2,E_0)\,$, whose spectra for $J^0$ are neither 
bounded from above nor from below, 
contain the principal and complementary (or supplementary) UIRs of 
$SO(2,1)\,$. As explained in the next Section, we shall  focus on the 
other possible UIR's. 
The representations ${\cal D}^{\pm}(\Phi)$ are called the 
\emph{discrete series}, 
while the representation ${\cal{D}}(0)$ is 
the trivial, one-dimensional representation. 
For the discrete series ${\cal D}^{+}(\Phi)$ 
(resp. ${\cal D}^{-}(\Phi)$), 
the spectrum of $J^0$ is countably infinite, 
bounded from below (resp. above). Cases of particular interest 
for our purpose will be denoted by 
\begin{eqnarray}\label{reps}
{\cal D}^{+}_{s}&:=&{\cal D}^+(-s)\ :\ J^0 = s + m\;, \nonumber \\
{\cal D}^{-}_{s}&:=&{\cal D}^-(-s)\ : \ J^0= -s - m\;,\quad 
m\in \mathbb{N}\;,\quad s > 0\;.
\end{eqnarray}
The spin of the discrete series ${\cal D}^+_{s}$ representation 
is $s\,$ (with $s>0\,$), that can be integer or even an 
arbitrary (albeit positive), real number. 

Representations of $SO(2,1)$  bounded from above and below like for 
$SO(3)$ exist but are non unitary \cite{Barut:1965}. 
The only UIR that $SO(3)$ and $SO(2,1)$ share 
is the trivial one ${\cal D}(\Phi=0)$, which corresponds to scalar 
fields. It is finally worth 
mentioning that the statistical phase $\exp{(2 i\pi s)}$ can still be 
associated with a state of arbitrary spin $s$ by virtue of the 
spin-statistics theorem \cite{Frohlich:1988qh,Mund}. 

\subsection{The case of ISO(2,1)}

Let $P^a$ be the translation generators of $T_3\,$. 
Then $iso(2,1)\,$, the Poincar\'e algebra in $2+1$ dimensions, is 
presented by 
\begin{equation}\label{alg0}
[J^a ,J^b]= - i \varepsilon^{abc} J_c\;, \quad 
[J^a, P^b]= -i \varepsilon^{abc} P_c\;, \quad 
[P^a ,P^b]=0\; ,
\end{equation}
and the two Casimir operators of $ISO(2,1)$ read, for massive 
representations,
\begin{equation}\label{casimir}
M^2=- {\bm P}^2, \quad s=-\frac{{\bm P}\cdot {\bm J}}{M}.
\end{equation}
They respectively give, on irreducible representations, the squared 
mass and the spin of a state. It has been shown in 
\cite{Cortes:1992fa} that states $\lvert\Psi\rangle$ belonging to the 
complementary series ${\cal D}(C_2,E_0)$ are such that 
${\bm P^2}\lvert\Psi\rangle=0$, ${\bm P}\cdot {\bm J}\lvert\Psi\rangle=0$, ${\bm J}^2\lvert\Psi\rangle=0$. 
Such states are not relevant in view of studying glueballs since we 
are looking for massive representations with nonzero spin, that will 
contain anyons. As also shown in \cite{Cortes:1992fa}, such 
``physical" states belong rather to the discrete series 
${\cal D}^+_s$ or ${\cal D}^-_s\,$. 
Let us denote by $\lvert M^2;\vec p ; s; J^0\rangle $ these states, 
the two series being distinguished by the signs of the eigenvalues of 
$P^0$ and $J^0$: positive (resp. negative) for ${\cal D}^+_s$ 
(resp. ${\cal D}^-_s$). 
Therefore, the two series ${\cal D}^\pm_s$ can be seen as 
${\cal PT}$ conjugated to each other, ${\cal P}$ and ${\cal T}$ denoted respectively the parity and time-conjugation. 
In the rest frame , $\vec p=\vec 0$, and $s$ reduces to $J^0$ 
(resp. $-J^0$) for states in the  ${\cal D}^+_s$ 
(resp. ${\cal D}^-_s$) representation. 
We note 
\begin{equation}\label{basis}
\lvert M^2 ; s;\pm s\rangle \in  {\cal D}^\pm_s 
\end{equation}
such states, that will play a particular role in the rest of this work. 

In $2+1$ dimensions, the action of parity ${\cal P}$ is to revert 
one spatial direction; we define it to act as 
${\bm X}=(x^0,x^1,x^2) \rightarrow  {\cal P}{\bm X}{\cal P}^{-1} = (x^0,x^1,-x^2)$. 
As a consequence, 
\begin{equation}\label{alg0b}
[{\cal P},{\bm P}^2]= 0\;, \quad 
\{ {\cal P}, \tfrac{-{\bm P}\cdot {\bm J}}{M} \}=0\, .
\end{equation}

Eigenstates of both (\ref{casimir}) 
and the parity can be built; 
they represent anyons and in the 
rest frame they read
\begin{equation}\label{doublet}
\lvert M^2 ; s; \eta_{\cal P}\rangle=\frac{1}{\sqrt 2} 
\left( \lvert M^2 ; s; s \rangle  
+ \eta_{\cal P} \lvert M^2 ; s;- s\rangle\right) 
\quad \in\quad {\cal D}^+_s\oplus {\cal D}^-_s ,
\end{equation}
where $\eta_{\cal P}$ is the eigenvalue of the parity. 
This prescription is valid when $s\neq 0$. For states belonging to 
${\cal D} (0)$, eigenstates of the parity can still be obtained by 
application of the projector $\frac{1}{2}(1+\eta_{\cal P} {\cal P})$, but both 
values of $\eta_{{\cal P}}$ cannot necessarily be reached, as we will see in Sec. \ref{secmodel} by explicit computation.

\section{Anyons from closed strings}\label{secexam}

As shown in \cite{Mezi}, fractional spin do appear in the 
spectrum of closed $(2+1)$-dimensional Nambu-goto strings
in the light-cone gauge. 
More precisely, the authors of \cite{Mezi}  
have performed the light-cone quantization of the following 
Hamiltonian version of the Nambu-Goto action 
\begin{equation}
S[{\bm X},{\bm P}; l, u]=\int d\tau \int \frac{d\phi}{2\pi} 
\left\lbrace  \dot {\bm X}\cdot {\bm P}-\frac{l}{2}\left[ 
{\bm P}^2+(2\pi\sigma {\bm X}')^2\right] -u {\bm X}'
\cdot{\bm P}\right\rbrace ,
\label{HamiltonAction}
\end{equation}
where $\sigma$ is the string tension and where the string 
coordinates ${\bm X}$ are a function of $\tau$ and 
$\phi\in[0,2\pi]$. 
This last action is equivalent to the standard Nambu-Goto 
action provided $l$, the Lagrange multiplier accounting 
for the S$^1$-diffeomorphism invariance, is nowhere 
vanishing. The other Lagrange multiplier, $u$, stands 
for the $\tau$-reparametrization invariance. 
The reader can find in \cite{Brink:1988nh}
a detailed and rigorous presentation of the Hamiltonian 
quantization of the Polyakov action for the 
(super)string, where the Hamiltonian action 
(\ref{HamiltonAction}) appears upon fixing the 
constraint related to Weyl invariance of the 
classical Polyakov string.

A first observation made in \cite{Mezi} is that the mass 
spectrum of the theory reads 
\begin{equation}\label{massfor}
M^2=4\pi\sigma(N+\bar N-a),
\end{equation}
with the usual number operators $N$ and $\bar N$.
The constraint
\begin{equation}
N=\bar N
\end{equation}
equating the number of left- and right-movers, as a 
consequence of the S$^1$-diffeomorphism invariance, 
must be added to Eq. (\ref{massfor}). The constant $a$ 
is actually not constrained by the theory. 
Indeed, it is well known that  a light-cone 
quantization in a $D$-dimensional spacetime would 
have led to the critical value $a=(D-2)/12$ necessary 
to restore the Lorentz invariance at quantum level. 
However the authors of \cite{Mezi} have fixed 
$D=3$ \textit{a priori}, which has a strong impact: The 
problematic commutators are \textit{de facto} absent 
and Poincar\'e invariance is satisfied at the quantum 
level without having to fix $a$ unless the theory is 
supersymmetric, a case that we are not dealing with 
here. 

The spectrum can be built by requiring the string 
states to be simultaneously eigenstates of $M^2$ and 
$s$, given by (\ref{casimir}). This last operator is 
cubic in the $\alpha'$s and couples the different 
states with the same $N$. The eigenvalues  of the
operator $s$ finally give the spins of the closed 
string states with a given mass. 
Inspection of these eigenvalues shows that there 
necessarily are fractional spin fields in the spectrum
of the first-quantized closed string in 3D. 
This is the key result of \cite{Mezi}. 
More precisely, the first 
levels of the closed string spectrum contain states 
with the following spins: 
\begin{itemize}
\item Only $s=0$ for $N=0$ and $N=1\,$; 
\item  Two $s=0$ states and two 
$ s  =\frac{3}{\sqrt{4-a}}$ 
states for $N=2\,$; 
\item  Three $s=0$ states, four 
$s  =\sqrt{\frac{179}{12\sqrt{6-a}}}$ states
and two $s =\sqrt{\frac{179}{3\sqrt{6-a}}}$ 
states for $N=3\,$.  
\end{itemize}
States with $s\neq 0$ actually appears in 
doublets of opposite helicities, standing for the two 
discrete series ${\cal D}^\pm_s\,$. 
We recall that both discrete series are 
characterised by the same eigenvalue of the 
the operator on the right-hand side of the
second equation of \eqref{casimir}, but differ 
by the sign of $J^0\,$. 
The interested reader will find the explicit 
expression of all the above states in terms of the 
string oscillators in Ref. \cite{Mezi}. 

There is actually an infinite but countable set of 
closed string states, some of which having 
fractional spin 
since there is no value of $a$ leading to only integer or 
half-integer spins. In view of what we recalled in 
Sec. \ref{secgen}, this result is natural: Imposing  
Poincar\'e invariance to the first-quantised closed 
string in 3D should logically lead to states belonging to 
anyonic representations. Note however that the 
non-critical nature of the bosonic string in 
$(2+1)$ dimensions comes in the light-cone 
quantisation prescription. BRST quantisation, 
on the other hand, 
forbids low-dimensional, critial  Polyakov strings, see 
\cite{Brink:1988nh}.

\section{The model}\label{secmodel}

\subsection{Glueballs and closed strings}

Beyond the pioneering work \cite{Vene}, the relevance of 
relating Yang-Mills theory at large-$N_c$ to a closed 
string theory has been studied also in \cite{Bars}, where 
the following picture is developed. 
On the one hand, at large $N_c$, Yang-Mills dynamics 
can be reformulated in terms of a reduced model, 
typically a quenched Eguchi--Kawai model \cite{QEK}. 
On the other hand, an appropriate limit 
$N_c\rightarrow \infty$ of $SU(N_c)$ 
\cite{Bars,Fairlie:1989vv} 
is isomorphic to the algebra of area-preserving 
diffeomorphisms. 
Both results allow to reformulate the quenched 
Eguchi--Kawai action as a Nambu--Goto action.
However, $SU(\infty)$ Yang-Mills is not fully equivalent 
to a Nambu--Goto string, since the integration measure of 
its partition function is not that of a Nambu-Goto string \cite{Bars}. 
Other approaches clearly show that a closed Nambu--Goto 
string can only be a leading-order  approximation of 
Yang-Mills theory even at large-$N_c$, see \textit{e.g.} 
\cite{Make} and references therein. 

An other point of view is the one of \cite{klebanov}, in which Yang-Mills theory in $2+1$ dimensions is reduced to a $(1+1)$-dimensional Yang-Mills theory with scalar adjoint matter. The spectrum of the latter theory is shown to contain bound states (glueballs) that can be interpreted as closed strings. Nevertheless, as observed in \cite{Bars}, the Nambu--Goto string alone 
cannot provide an effective description of Yang-Mills 
theory. A better-known reason is the standard result that 
Poincar\'e invariance is fulfilled at the quantum level for 
$D=26$ only. This issue was solved in \cite{PS}, where 
it was shown that adding a term
\begin{equation}
\delta {\cal L}_{PS}\propto 
\frac{(\partial_\alpha \partial_\beta X^\mu 
\partial^\beta X_\mu)^2}{(\partial_\gamma X_\mu)^2}
\end{equation}
to the Polyakov Lagrangian restores Poincar\'e invariance 
for any spacetime dimension $D\,$. The  
Polchinski--Strominger term has been computed in conformal 
gauge \cite{PS} and recovered in static gauge \cite{PS2}. 
Note that such an extra term is not needed in the case 
we focus on since, within the light-cone gauge quantisation 
scheme used in \cite{Mezi}, Poincar\'e invariance is already 
satisfied at the quantum level for the 3D Nambu--Goto action.

Another reason to go beyond the Nambu--Goto string may then be 
to reach a more accurate description of the dynamics of the 
effective QCD string. For example, as seen from a 
semiclassical expansion around a closed folded string, the 
Polchinski--Strominger term produces corrections to the 
well-known mass formula $M^2\propto J\,$, $J$ being the string 
angular momentum. The corrections appear as powers of $J$ 
smaller than one and have been computed in \cite{regge}. 
More generally, the analysis performed in \cite{aha} of the 
terms allowed by classical Lorentz invariance reveals that the 
first nontrivial correction to the Nambu--Goto Lagrangian in 
$2+1$ dimensions is a term involving the induced worldsheet 
metric $h$ and the scalar curvature $R$ constructed from it:
 \begin{equation}
\delta {\cal L}\propto \sqrt{-h} \,R^2 .
\end{equation}
However, in the present exploratory work, we are mainly 
interested in a qualitative description of the glueball 
spectrum, so it is worth asking whether adding such a term 
brings relevant information or not. 
It appears from Ref. \cite{aha2} that, expanding the energy of 
an effective closed string in terms of its classical length 
$L\,$, the energy formula is universal up to $1/L^5$ terms  
in $2+1$ dimensions and deviations from universality only 
appear at order $1/L^7\,$.  
According to lattice computations \cite{Teper2}, 
the mass of the lowest-lying glueball at large$\,$-$N_c$ is 
given 
by $M/\sqrt\sigma\sim 4\,$, which provides the estimate 
$ \sqrt \sigma L\sim 4\,$, a length range such that 
$1/(\sqrt{\sigma} L)^7\,$ corrections to the standard 
Nambu--Goto 
energy formula are negligible \cite{athe}.

We aim at building an effective model in which the 
nonperturbative dynamics of $(2+1)$-dimensional YM theory 
is that of a closed bosonic string. From what we have just 
been arguing, it is thus sufficient to adopt, in a first 
approach, the quantization scheme of \cite{Mezi} that will 
allow us to reach this goal. 

\subsection{Glueball states}

In order to match string states and glueball states according 
to standard terminology, one has to associate $s^{{\cal PC}}$ 
quantum numbers to a given string state. On top of the reversal of 
any spatial momentum, the parity operator ${\cal P}$ for closed 
strings is defined by 
\begin{equation}
{\cal P}= (-1)^{N+\bar N}.
\end{equation}
It anticommutes with the helicity operator \cite{Mezi}. 
As a consequence, for any given eigenvalue of $N$ 
(equivalently $M^2$), states with nonzero spin form parity 
doublets (\ref{doublet}). The $s=0$ cases must be treated 
separately, see below. 

Charge conjugation ${\cal C}$ has to be introduced by hand by 
recalling that, in $2+1$ dimensions, a closed flux tube is actually a loop of fundamental color flux that closes 
on itself. Hence it has an intrinsic orientation which is that of the chromoelectric 
field \cite{John}. 
So a given state in the closed-string spectrum can 
either correspond to a flux tube with clockwise ($\rightturn$) 
orientation or anticlockwise orientation ($\leftturn$). 
The action of the charge conjugation is to revert this 
orientation, basically by turning fundamental color charges into 
conjugated ones \cite{John},
\begin{equation}
{\cal C}\lvert \rightturn; M^2; s; s\rangle = 
\lvert \leftturn; M^2; s; s\rangle,
\end{equation}
while parity also flips $J^0\,$:
\begin{equation}
{\cal P}\lvert \rightturn; M^2; s; s\rangle=\lvert 
\leftturn; M^2; s; -s\rangle.
\end{equation}
Note that, in our framework, time reversal would just flip 
$J^0\,$. 

In summary, starting from a closed-string state 
$\lvert\rightturn; M^2; s; s\rangle$ found in \cite{Mezi}, 
one can build a $s^{\eta_{\cal P}\eta_{\cal C}}$ glueball with 
mass $M^2$ provided that the linear combination
\begin{equation}\label{JPCdef}
\lvert M^2; s^{\eta_{\cal P}\eta_{\cal C}}\rangle = 
\tfrac{1}{2}\,(1+\eta_{\cal C} 
{\cal C})(1+\eta_{\cal P} {\cal P})
\lvert \rightturn; M^2; s; s\rangle
\end{equation}
is nonzero. 
At this stage, charge conjugation just adds an additional 
$\mathbb{Z}_2$ degree of freedom 
to the spectrum.

The explicit form of the eigenstates of $M^2$ and $s$ 
is given in \cite{Mezi} and will not be recalled here 
for the sake of brevity. 
We have checked that, from these 
$\lvert\rightturn; M^2; s;s\rangle$ states, one can form the 
following multiplets:
\begin{itemize}
\item $\{0^{++},\ 0^{--}\}$ for $N=0$;
\item  $\{0^{++*},\ 0^{--*}\}$ for $N=1$;
\item $\{0^{++**},\ 0^{-+},\ 0^{--**},\ 0^{+-},\ \frac{3}{\sqrt{4-a}}^{\pm\pm}\}$ for $N=2$. 
\item \dots
\end{itemize}
The $^*$ is used to distinguish 
excited states of a given $s^{\eta_{\cal P}\eta_{\cal C}}\,$. 
It is readily seen that, if glueball dynamics is that of a closed 
string, the low-lying spectrum should be filled by (pseudo)scalar 
states, while the first states with nonzero spin are expected to 
arise at higher masses, corresponding to level $2$ in our 
formalism. At this stage, the state with $s=3/\sqrt{4-a}$ can 
still be a boson with spin $n\in\mathbb{N}_0$ 
provided that $a=4-9/n^2\,$. 
However, $n>1$ leads to $a>0\,$, implying 
unphysical glueball states with $M^2<0$ at level 0. 
Even if the $N=2$ glueball with $J\neq 0$ is not an anyon but a 
spin-$1$ boson, then anyons necessarily appear at level $3$, so 
they cannot be avoided in the glueball spectrum. 

\section{Glueball spectrum}\label{secresu}

\subsection{Numerical results}

Glueball states obtained in the previous section follow the 
simple mass formula (\ref{massfor}). Hence the glueball spectrum 
is completely known from our model once the value of $a$ is 
fixed. 
As usually done in the field, this can be achieved by comparing 
our results to the $(2+1)$-dimensional glueball spectrum computed 
in pure gauge lattice QCD in Refs. \cite{Teper,Teper2} 
and further analyzed in \cite{Teper3,Teper4}. 

A clear feature of the lattice spectrum is the appearance of 
Regge trajectories, \textit{i.e.} a linear dependence between 
the squared mass $M^2$ and the spin $s$ of a glueball, 
with a slope compatible with the value $8\pi\sigma$ of a 
classical closed string \cite{Teper3,Teper4}. 
However the spin ``measured" on the lattice is necessarily 
integer due to the Euclidean spacetime induced by Wick's 
rotation. That is why, as discussed in Sec. \ref{secgen}, 
comparisons between our model and lattice results should be 
restricted to $s=0$ states: $SO(2,1)$ and $SO(3)$ only share the 
${\cal D}(0)$ UIR. 
These states are listed in Table \ref{tab}. It is readily seen 
that, as predicted by the closed-string picture, the lightest 
states with $C=+$ (resp. $C=-$) are $0^{++}$ (resp. $0^{--}$) 
ones, while the first $0^{-+}$ (resp. $0^{+-}$) glueball is 
much heavier. 

As pointed out in \cite{John}, the lattice spectrum 
shows a large splitting between $C=+$ and $C=-$ states, which are 
degenerate according to the mass formula (\ref{massfor}). 
As argued in \cite{John}, this is the stage at which it has to be 
remembered that flux tubes may be more complex objects than 
Nambu--Goto strings because of their intrinsic orientation. 
Processes that induce a mixing between $\rightturn$ and 
$\leftturn$ states can be figured out: One can think of a 
$\rightturn$ flux tube shrinking to a ``ball-like" configuration 
where information about the orientation is lost, then expanding 
into a $\leftturn$ flux tube. The simplest way of implementing 
such a mixing is to add a constant coupling of the form 
\begin{equation}
\begin{pmatrix}
M^2 & 4\pi\sigma b  \\ 
4 \pi\sigma b & M^2 
\end{pmatrix} \;,
\end{equation}
the above mass (squared) operator being expressed in the  
$\{\lvert\rightturn\rangle, \lvert\leftturn\rangle\}$ basis. 
The eigenstates are $C=+$ states, with mass 
$M^2_{C=+}=4\pi \sigma (N+\bar N-a-b)$ and the $C=-$ states, 
with mass $M^2_{C=+}=4\pi \sigma (N+\bar N-a+b)$. 
The effect of the mixing introduced is thus simply to shift the 
intercept of $C=-$ states with respect to that of $C=+$ states.

\begin{table}[b]
\begin{tabular}{c|lcr|lcr}

&  &  \multicolumn{2}{c|}{$M/\sqrt\sigma$}  &   &   \multicolumn{2}{c}{$M/\sqrt\sigma$} \\
\hline
$N$ & $s^{\eta_{\cal P}\eta_{\cal C}}$ &  Model & Lattice  & $s^{\eta_{\cal P}\eta_{\cal C}}$  &  Model & Lattice  \\
\hline 
0 & $0^{++}$ & 4.081 & 4.108(20) \cite{Teper2} & $0^{--}$ & 5.950 & 5.953(71) \cite{Teper2}\\
\hline 
1 &$0^{++*}$  & 6.464 & 6.211(46) \cite{Teper2} &$0^{--*}$  & 7.780 & 7.77(14) \cite{Teper2}\\ 
\hline 
2 &$0^{++**}$ & 8.180 & 8.35(20) \cite{Teper2} &$0^{--**}$ & 9.256 & 8.96(65) \cite{Teper} \\ 
   &   $0^{-+}$  & 8.180  & 9.02(30) \cite{Teper2}&$0^{+-}$ & 9.256 & 9.47(116) \cite{Teper} \\
   & $1.22^{\pm+}$ & 8.180   && $1.22^{\pm-}$ & 9.256  &  \\
\end{tabular}  

\caption{Glueball quantum numbers predicted by our flux tube 
model, with $a=-2.071$ and $b=0.746$, compared to the pure gauge 
lattice studies \cite{Teper,Teper2} in the large$-N_c$ limit. 
Masses are given in units of the string tension.}
\label{tab}
\end{table}

The model built here is obviously very simple and should be 
regarded as valid only in a first approximation. 
Spin-dependent corrections, in particular, should be present 
in a more refined model. It is nevertheless interesting to notice the good agreement between our mass formula and existing lattice 
data once $a$ and $b$ are fitted, see Table \ref{tab}. 
A prediction of the present model is that there should exist two 
degenerate $1.218^{\pm+}$ glueballs with a mass around $8.18$ in 
units of the string tension, as well as $1.22^{\pm-}$ glueballs 
with a mass around $9.26\,$. 

For completeness we mention that an attempt to compute the 
large-$N_c$ glueball spectrum in $(2+1)$ dimensions by resorting 
to a formulation of lattice gauge theory in the light-cone gauge 
has been made previously \cite{Dalley:2000ye}. 
Among other results the ratios $M_{0^{--}}/M_{0^{++}}=1.35(5)$ 
and $M_{0^{--*}}/M_{0^{++}}=1.82(6)$ are found, while our 
approach leads to the similar values 1.46 and 1.90 respectively, 
keeping the same values of $a$ and $b\,$. 
Anyonic states were not built in Ref. \cite{Dalley:2000ye}; 
to our knowledge it is an open question to know whether anyonic states can be built in light-cone gauge lattice theory or not. 

\subsection{Comments on the mass spectrum}

Although the present flux tube model is close to the one proposed 
in \cite{John}, a fundamental difference occurs at the level of 
the quantization of the closed string. Indeed, in \cite{John}, 
a spectrum was found in agreement with lattice data by using the 
Isgur--Paton closed flux-tube model \cite{Isgu}. This is not 
surprising since the authors of \cite{John} perform a 
nonrelativistic, Schr\"odinger-like quantisation of the 
fluctuations of a closed circular string, and in such a scheme 
the spin of a state is identified with $s=\vert N- \bar N\vert\,$ 
so it is necessarily an integer and the constraint $N=\bar N$ is 
not present. Only the constraint $N+\bar N\neq 1$ is imposed by 
the model \cite{John}. Hence, the angular momentum appearing in 
the resulting Hamiltonian is integer and matches existing lattice 
data.

When $N_c$ is finite, our main assumption --- \emph{i.e.} 
identifying glueballs with closed flux tubes --- may appear less 
sound. It has to be noticed however that the quantum numbers and 
mass hierarchy of the glueball states are identical whatever 
$N_c$ is \cite{Teper,Teper2}. The case $N_c=2$ is 
special since the fundamental representation is real. 
Then, no orientation can be given to a flux tube, and only the 
$C=+$ sector is present. The universal structure of the glueball 
spectrum for $N_c>2$
may suggest that the stringy picture developed here is still 
relevant at finite $N_c$ and thus that anyonic glueballs are a 
generic feature of $SU(N_c)$ Yang-Mills theory in 
$(2+1)$ dimensions. Even the $SU(2)$ lattice scalar mass spectrum can be recovered by using $b=0$ (no $C=-$ sector) and $a=-1.9$ in our model. Note that the spectrum obtained in the present section is 
expected to be the same in the large $N_c$ limit of $SU(N_c)\,$, 
$SO(N_c)$ and $Sp(N_c)$ Yang-Mills theories, 
that have been proven to be equivalent in the strong coupling 
limit \cite{Lovelace:1982hz}.  

\section{Relation with 't Hooft and Wilson 
loops}\label{secrel}

It is now worth wondering how much the existence of 
anyonic states in YM theory relies on our effective 
closed-string description. 
There exist other ways to build anyons.  
One of the simplest ways, at the nonrelativistic 
level, is to minimally couple a particle to a 
vortex-like vector potential: The resulting 
vortex-plus-particle system constitutes an anyon \cite{Fort}. 
This coupling can be achieved in Yang-Mills theory too. 
Let us start from the 3D 't Hooft operator $\phi(\vec x)$ 
defined through the nonstandard commutation relation \cite{'tHooft} 
\begin{equation}
W(C_{t})\phi(\vec x)
={\rm e}^{\frac{2\pi i n(\vec x;C_{t})}{N_c}}\phi(\vec x)W(C_{t}),
\end{equation}
where  $W(C_{t})={\rm Tr\, P\, exp\, }ig\oint_{C_{t}} A\;$
is a standard Wilson loop with $C_{t}$ a closed spacelike curve. 
By ``spacelike" it is meant  that all the points of $C_{t}$ have 
the same temporal coordinate $x^0=t\,$. 
Moreover, in the equation above, $n(\vec x;C_t)$ is the number of 
times that the closed curve $C_{t}$ winds around $\vec x$ in a 
clockwise fashion minus the number of times it 
winds around $\vec x$ anticlockwise. 
Note also that $[\phi(\vec x),\phi(\vec y)]=0$ which reflects
the locality of the operator $\phi\,$ \cite{'tHooft}. Explicit representations 
of $\phi(\vec z)$ can be found in \cite{Itzhaki,Rein}. 

We now define the operator 
\begin{equation}\label{Gdef}
G_{C_t}(\vec z)=\phi(\vec z)W(C_{t}),
\end{equation}
where  $\vec z$ may or may not be enclosed by $C_t$, a closed 
spacelike curve fixed once for all. Since spacelike Wilson loops 
commute at equal time \cite{'tHooft}, it is readily shown that 
$G_{C_t}(\vec z)$ 
may have a nontrivial statistical phase: It is indeed such that, 
for two separated points $\vec z_1$ an $\vec z_2\,$,
\begin{equation}
G_{C_t}(\vec z_1)G_{C_t}(\vec z_2)
={\rm e}^{\frac{2\pi i}{N_c}[ n(\vec z_2;C_t) 
- n(\vec z_1;C_t)]}G_{C_t}(\vec z_2)G_{C_t}(\vec z_1)\;.
\end{equation}
The statistical phase will be nontrivial as soon as 
$\,n(\vec z_2;C_t)\neq n(\vec z_1;C_t)\,$. 
From the generalized spin-statistics theorem \cite{Mund}, 
it can be concluded that the operator $G_{C_t}(\vec z)$ creates a 
color-singlet state with spin $s = (k/N_c)+n\,$ with 
$k,n\in\mathbb{N}\,$, 
that is, a value that can be nonzero and neither integer nor 
half-integer. 

Just as the correlator of spacelike Wilson loops contains scalar 
glueballs \cite{Teper}, it can be expected that the correlator 
 $\langle 0 \rvert G^\dagger_{C_t}(\vec z) G_{C_0}(\vec z)\lvert 0 \rangle$
will propagate anyonic glueballs with spin $k/N_c\,$. If that turned out to be true, this would show 
that our main result is not fully dependent on the model used. 
In the context of the Abelian Higgs model with Chern-Simons term, 
the propagation of anyonic states is described in 
\cite{Frohlich:1988qh}, 
where in particular it is shown that the 
physical Hilbert space of 1-$\,$anyon states is decomposed into 
orthogonal sectors labelled by the vorticity $q\,$: 
\begin{equation}
{\cal H}^{(\mu)} = 
\bigoplus_{q \in \mathbb{Z}}{\cal H}_q^{(\mu)}\;,
\end{equation}
where $\mu/4\pi$ is the coefficient multiplying the Chern-Simons
term $\int A\wedge dA\,$ in the action and  where 
the vorticity eigenvalue $q$ labels the homotopy classes for the 
map $\mathbf{S}^1 \rightarrow \mathbf{S}^1$ expressing 
the asymptotic behaviour of the complex scalar field at 
spatial infinity. 
Note that the spin of a state is then given by 
$\mu q^2/2 $ mod $\mathbb Z\,$.
The previous considerations on anyon propagation 
can even be made more rigorous on the 
lattice in 3D Euclidean space, 
see Sec. 7 of \cite{Frohlich:1988qh}. 

\section{Summary and outlook}\label{secconclu}

In this note, we have developed a closed-string model of glueballs in $(2+1)$ 
dimensions based on the light-cone quantization of the 
Nambu--Goto string performed in \cite{Mezi}. 
Since closed-string are actually used to model the dynamics of 
Yang-Mills field, the orientation has been added as an extra 
quantum number in order to account for the fact that we are 
dealing with effective rather fundamental strings. 
This addition has two consequences: The possibility of defining 
the charge-conjugation of a state, and the addition of a mixing 
mechanism eventually splitting the masses of states with 
different eigenvalues under charge conjugation. 
Our model has two free parameters that, once fitted, allow to 
satisfactorily reproduce the masses of the 8 zero-spin 
glueballs currently observed in large-$N_c$ lattice calculations. 
As a consequence of our model, anyonic glueballs must be present 
with a mass and spin that both depend on the intercept 
$\left. \frac{M^2}{4\pi\sigma}\right|_{N=\bar N=0}\,$. 

We believe that the existence of such states is not an 
artifact of the closed-string picture proposed, but rather, 
that it is a generic property of Yang-Mills theory in 
$(2+1)$ dimensions. 
Hence, the existence of anyonic glueballs could be confirmed 
(or not) in the future by resorting to lattice calculations, 
either in light-cone gauge or in the more standard temporal gauge 
provided that appropriate correlators are built. As a starting 
point for future calculations, an inspiring explicit form for the t'Hooft 
operator can be 
found in \cite{Rein}, while similar results have been proposed 
in the framework of the Abelian Higgs model in \cite{akh}.

\section*{Acknowledgements}

We thank Th. Basile, M. Chernodub, J. Evslin, M. 
Henneaux, M. Porrati, P. Sundell and M. Valenzuela for 
discussions. N.B. is F.R.S.-FNRS Research Associate 
(Belgium) and his work was supported in parts by an
ARC contract No. AUWB-2010-10/15-UMONS-1.

\end{document}